\documentstyle[twocolumn,prl,aps]{revtex}

\begin{document}

\relax 

\draft \preprint{FM 96-2} 

\title{Extension of Onsager's reciprocity to large fields and the 
chaotic hypothesis}

\author{Giovanni Gallavotti} 
\address{ Fisica, Universit\'{a} di Roma, ``La Sapienza'', 00185
Roma, Italia}

\date{\today}
\maketitle
\begin{abstract}
The fluctuation theorem (FT), the first derived consequence of the {\it
Chaotic Hypothesis} (CH) of \cite{[GC1]}, can be considered as an extension
to arbitrary forcing fields of the fluctuation dissipation theorem
(FD) and the corresponding Onsager reciprocity (OR), in a class of
reversible nonequilibrium statistical mechanical systems.

\end{abstract}

\narrowtext

A typical system studied here will be $N$ point particles subject
to {\it (a)} mutual and external conservative forces with potential
$V(\vec q_1,\ldots,\vec q_N)$, {\it (b)} external (non conservative)
forces, {\it forcing agents}, $\{\vec F_j\}$, $j=1,\ldots,N$, whose
strength is measured by parameters $\{G_j\}$, $j=1,\ldots,s$, and {\it (c)}
also to forces $\{\vec\varphi_j\}$, $j=1,\ldots N$,
generating constraints that provides a model for the thermostatting
mechanism that keeps the energy of the system from growing indefinitely
(because of the continuing action of the forcing agents).

An observable $O(\{\vec q,\dot{\vec  q}\})$ evolves in time under the time 
evolution $S_t$ solving the equations of motion:

\begin{equation}{m\ddot{{\vec q}}}_j = 
-\partial_{\vec q_j} V({{\vec q}}_j)+ \vec F_j(\{G\})+\vec\varphi_j
\label{1.1}\end{equation}

\noindent{}with $m=$ particles mass, and $\vec \varphi_j$
"thermostatting" forces assuring that the system reaches a (non
equilibrium) stationary state.  The time evolution of the observable
$O$ on the motion with initial data $x=(\vec q,\dot{\vec q})$ is the
function $t\to O(S_t x)$ so that the motion {\it statistics} is
the probability distribution $\mu_+$ on the {\it phase space} ${{\cal
C}}$ such that:

\begin{equation}{\lim_{T\to\infty}{1\over T}
\int_0^T O(S_tx) \,dt=\int_{{\cal C}} O(y)\mu_+(dy)}
\,{\buildrel def\over =}\,\langle O\rangle_+
\label{1.2}\end{equation}

\noindent{}for all data $x\in{{\cal C}}$ except a set of zero measure 
with respect to the volume $\overline\mu_0$ on ${{\cal C}}$. The
distribution $\mu_+$ is assumed to exist: a property called {\it zero-th
law}, \cite{[GC2],[G4]}. 

The thermostatting mechanism will be described by force laws
$\vec\varphi_j$ which enforce the constraint that the kinetic energy (or
the total energy) of the particles, or of subgroups of the particles,
remains constant, \cite{[1]}.  It is convenient also to imagine that the
constraints keep the total kinetic energy bounded (hence phase space is
bounded). The constraint forces $\{\vec\varphi_j\}$ will be supposed
such that the system is {\it reversible}: this means that there will be
a map $i$, defined on phase space, anticommuting with time evolution:
{\it i.e.}  $S_t\,i\equiv i\,S_{-t}$.

Reversibility is a key assumption, \cite{[GC1],[R]}.

In \cite{[GC2]} the divergence of the r.h.s. of Eq. (\ref{1.1})
is a quantity $-\sigma(x)$ defined on phase space that has been 
identified with the {\it entropy production rate}.

The {\it chaotic hypothesis}, (CH), of \cite{[GC1]} implies a {\it
fluctuation theorem} or (FT) which, \cite{[GC2]}, is a property of the
fluctuations of the entropy production rate.  Namely if we denote
$\langle\sigma\rangle_+=\int_{{\cal C}} \sigma(y)\mu_+(dy)$ the time
average, over an infinite time interval by Eq.  (\ref{1.2}), then the
{\it dimensionless} finite time average $p=p(x)$:

\begin{equation}{{1\over \tau}
\int_{-\tau/2}^{\tau/2}
\sigma(S_tx)\,dt\,{\buildrel def\over =}
\,\langle\sigma\rangle_+\,p}
\label{1.3}\end{equation}

\noindent{}has a statistical distribution $\pi_\tau(p)$
with respect to the stationary state
distribution $\mu_+$ such that:

\begin{equation}{{1\over \tau \langle\sigma\rangle_+\,p}
\log{\pi_\tau(p)\over\pi_\tau(-p)} \to_{\tau\to\infty}\,1}
\label{1.4}\end{equation}

\noindent{}provided (of course) $\langle\sigma\rangle_+>0$. Following
\cite{[GC1]}  a reversible system for which $\langle\sigma\rangle_+>0$
will be called {\it dissipative}.

Ruelle's $H$-theorem, \cite{[R]}, states that $
\langle\sigma\rangle_+>0$ unless the stationary distribution $\mu_+$
has the form $\rho(x)\overline\mu_0(dx)$. 

Hence we shall suppose that the system is dissipative when the forcing
$\vec G$ does not vanish and, without real loss of generality, that
$\sigma(ix)=-\sigma(x)$ and that $\sigma(x)\equiv0$ when the external
forcing vanishes, writing:

\begin{equation}
\sigma(x)=\sum_{i=1}^s G_i J^0_i(x)+ O(G^2)\label{1.4a}\end{equation}

Assuming that the $\sigma$--$\sigma$ correlations have a fast decay (a
fact to be expected if the (CH) is accepted, \cite{[S]})
then by a result of Sinai \cite{[S]} the entropy
verifies a {\it limit theorem}; {\it i.e.}:

\begin{equation}\lim_{\tau\to\infty}
{1\over\tau}\log\pi_\tau(p)=-\zeta(p)\label{1.5}\end{equation}

\noindent{}where 
$\zeta(p)$ is analytic for $p$ in the interval $[-p^*,p^*]$ within which
it can vary (model dependent)\cite{[G1],[Ge]}.  The function $\zeta(p)$
can be conveniently computed because its transform
$\lambda(\beta)=\lim_{\tau\to\infty}{1\over\tau}\log\int e^{\beta\tau\,
(p-1) \langle\sigma\rangle_+} \pi_\tau(p)dp$ can be expressed by a {\it
cumulant expansion}. Once $\lambda(\beta)$ is "known" then $\zeta(p)$ is
recovered via a Legendre transform; $\zeta(p)=
\max_\beta\big(\beta\langle\sigma\rangle_+(p-1)-\lambda(\beta)\big)
$, \cite{[G1],[Ge]}.

By using the cumulant expansion for $\lambda(\beta)$ we find that
$\lambda(\beta)={1\over 2!}\beta^2 C_2+{1\over3!}\beta^3 C_3+\ldots$
where the coefficients $C_j$ are
$\int_{-\infty}^\infty\langle\sigma(S_{t_1}\cdot)\sigma(S_{t_2}\cdot)\ldots
\sigma(S_{t_{j-1}}\cdot)\sigma(\cdot)\rangle^T_+\,dt_1\ldots$ if
$\langle\ldots\rangle^T_+$ denote the cumulants of the variables
$\sigma(x)$.

In our case the cumulants of order $j$ have size $O(G^j)$, by
Eq. (\ref{1.4a}), so that:

\begin{equation}
{\zeta(p)={\langle\sigma\rangle_+^2\over2 C_2}(p-1)^2+ O((p-1)^3
\big(G^3)}\label{1.6}\end{equation}

\noindent{}(note the first term in r.h.s. giving the central limit theorem).
Eq. (\ref{1.6}), together with the (FT) (\ref{1.4}),
yields at fixed $p$ our main relations:

\begin{equation}
\langle\sigma\rangle_+={1\over 2}C_2+ O(G^3)
\label{1.7}\end{equation}

We define, \cite{[G1]}, the {\it current} 
$J_i(x)=\partial_{G_i}\sigma(x)$ and 
the {\it transport coefficients} $L_{ij}=\partial_j\langle 
J_i(x)\rangle_+|_{G=0}$ and we study 
$L_{ij}$.

To derive (FD), we first look at the r.h.s. of the first relation in
Eq. (\ref{1.7}) discarding $O(G^3)$: the r.h.s becomes a quadratic form
in $G$ with coefficients:

\begin{equation}{
{1\over 2}\int_{-\infty}^\infty 
dt\, \big(\langle J^0_i(S_t\cdot) J^0_j(\cdot)\rangle_+-
\langle J^0_i\rangle_+\langle J^0_j\rangle_+
\big)}\label{1.8}\end{equation}

On the other hand the expansion of $\langle\sigma\rangle_+$ 
in the l.h.s. of 
Eq. (\ref{1.7}) to second order in $G$ gives:

\begin{equation}\langle\sigma\rangle_+={1\over2}
\sum_{ij} G_i G_j \big(\partial_i\partial_j\langle\sigma\rangle_+
\big|_{G=0}\big)\label{1.9}\end{equation}

\noindent{}because the first order term vanishes (by Eq. (\ref{1.4a}), or
(\ref{1.7})). The r.h.s. of (\ref{1.9}) is the sum of ${1\over2}G_iG_j$
times $\partial_j\partial_j \int \sigma(x) \mu_+(dx)$ which equals the
sum of the following three terms: the first is $\int\partial_i
\partial_j\sigma(x) \mu_+(dx)$, the second is
$\int\partial_i\sigma(x)\partial_j\mu_+(dx)+
(i\leftrightarrow j)$ and the third is
$\int\sigma(x)\partial_i\partial_j\mu_+(dx))$, all evaluated at
$G=0$. The first addend is $0$ (by time reversal), the third addend is
also $0$ (as $\sigma=0$ at $G=0$). Hence:

\begin{equation}\partial_i\partial_j \langle\sigma\rangle_+|_{G=0}=
\Big(\partial_j\langle{J^0_i}\rangle_++
\partial_i\langle{J^0_j}\rangle_+\Big)|_{G=0}
\label{1.10}\end{equation}

\noindent{}and it is easy to check, again by using time reversal, 
that:

\begin{equation}
\partial_j\langle{J^0_i}\rangle_+|_{G=0}\,=
\,\partial_j\langle{J_i}\rangle_+|_{G=0}
= \,L_{ij}\label{1.11}\end{equation}

\noindent{} Therefore equating the r.h.s and the l.h.s.
Eq.  (\ref{1.7}), as expressed respectively by
Eq.  (\ref{1.8}) and (\ref{1.10}) we get a formula for the matrix
${L_{ij}+L_{ji}\over2}$ giving (FD) ({\it i.e.} Green--Kubo's formula)
at least if $i=j$.  

We want to show that the above ideas {\it also} suffice to 
prove (OR), {\it i.e.} $L_{ij}=L_{ji}$. The main remark is that we can {\it 
extend} the (FT) theorem to give properties of the {\it joint} 
distribution of the average of $\sigma$, (\ref{1.3}), and of the 
corresponding average of $G_j\partial_j \sigma$. In fact define the 
{\it dimensionless } $j$-current $q=q(x)$ as:

\begin{equation}
{1\over\tau}
\int_{-\tau/2}^{\tau/2}
G_j\partial_j \sigma(S_tx)dt{\buildrel def\over =}
G_j\langle \partial_j\sigma\rangle_+\, q\label{1.13}\end{equation}

\noindent{}where the factor $G_j$ is there only to keep $\sigma$ and 
$G_j\partial_j\sigma$ with the same dimensions.

Then if $\pi_\tau(p,q)$ is the joint probability of $p,q$ the {\it same} 
proof of the (FT) in \cite{[GC1]} yields also:

\begin{equation}\lim_{\tau\to\infty} {1\over \tau \langle\sigma\rangle_+ p}
\log
{\pi_\tau(p,q)\over\pi_\tau(-p,-q)}=1\label{1.14}\end{equation}

\noindent{}and the limit theorem in (\ref{1.6}) is extended, \cite{[S]},
to:

\begin{equation}
\zeta(p,q)=\lim_{\tau\to\infty} 
{1\over\tau}\log\pi_\tau(p,q)\label{1.15}\end{equation}

We can compute $\zeta(p,q)$ in the same way as $\zeta(p)$ 
by considering first the transform $\lambda(\beta_1,\beta_2)$:

\begin{equation}
\lim_{\tau\to\infty}
{1\over\tau}\int e^{\tau(\beta_1\, (p-1)
\langle\sigma\rangle_++
\beta_2\,(q-1)\langle G_j\partial_j\sigma\rangle_+)}\pi_\tau(p,q)dpdq
\label{1.16}\end{equation}

\noindent{}and then the Legendre trasform:

$$\zeta(p,q)=\max_{\beta_1,\beta_2}\big(
\beta_1\, (p-1)\langle\sigma\rangle_++
\beta_2\,(q-1)
\langle G_j\partial_j\sigma\rangle_+-$$
\begin{equation}-\lambda(\beta_1,\beta_2)\big)\label{1.17}
\end{equation}

The function $\lambda(\vec\beta)$, $\vec\beta=(\beta_1,\beta_2)$, is 
evaluated by the cumulant expansion, as above, and one finds:

\begin{equation}
\lambda(\vec\beta)={1\over2}\,\big(\vec\beta, 
C\,\vec\beta)+O(G^3)\label{1.18}\end{equation}

\noindent{}where $C$ is the $2\times2$ matrix of the second order 
cumulants. The coefficient $C_{11}$ is given by $C_2$ appearing in 
(\ref{1.6}); $C_{22}$ is given by the same expression with $\sigma$ 
replaced by $G_j\partial_j\sigma$ while $C_{12}$ is the mixed cumulant:

\begin{equation}
\int_{-\infty}^\infty
\big(\langle\sigma(S_t\cdot)\,G_j\partial_j\sigma(\cdot)\rangle_+
-\langle\sigma(S_t\cdot)\rangle_+\,\langle
G_j\partial_j\sigma(\cdot)\rangle_+\big)dt\label{1.19}\end{equation}

Hence if $\vec w=\pmatrix{(p-1)\langle\sigma\rangle_+\cr
(q-1)\langle G_j\partial_j\sigma\rangle_+\cr}$ we get:

\begin{equation}
\zeta(p,q)={1\over2}\,\big(C^{-1}\vec w,\vec w\big)+O(G^3)
\label{1.20}\end{equation}

\noindent{}completely analogous to (\ref{1.6}). But the (FT) in 
(\ref{1.14}), implies that $\zeta(p,q)-\zeta(-p,-q)$
is {\it $q$ independent}: this means, as it is immediate to check:

\begin{equation} (C^{-1})_{22}
\langle G_j\partial_j\sigma\rangle_+-(C^{-1})_{21}
\langle \sigma\rangle_+=0+O(G^3)\label{1.21}\end{equation}

\noindent{}which because of (\ref{1.7}), and of 
$(C^{-1})_{22}=C_{11}/\det C$, becomes the analogue of (\ref{1.7}):

\begin{equation}
\langle G_j\partial_j\sigma\rangle_+={1\over2}C_{12}+O(G^3)
\label{1.22}
\end{equation}

\noindent{}and, proceeding as in the derivation of (\ref{1.8}) through 
(\ref{1.11}) ({\it i.e.} expanding both sides of (\ref{1.22}) {\it to 
first order} in the $G_i$'s and using (\ref{1.19})) we get
that $\partial_i\langle\partial_j\sigma\rangle_+$ is given by 
the integral in (\ref{1.8}). This means that $L_{ij}=L_{ji}$
and the (FD) theorem follows toghetr with the (OR).

Thus Eq.  (\ref{1.7}),(\ref{1.22}) and the ensuing (FD), and (OR), are a
consequence of (FT), (\ref{1.4}), and of its (obvious) extension,
(\ref{1.14}), in the limit $G\to0$, when combined with the expansion
(\ref{1.6}) for entropy fluctuations.  Those theorems and the fast
decay of the $\sigma\-\sigma$ correlations are all natural consequences
of (CH) for reversible statistical mechanical systems, which is the
starting point of our considerations.  Note that reversibility is here
assumed {\it both in equilibrium and in non equilibrium}: this is a
feature of gaussian thermostat models, \cite{[GC2]}, but by no means of
all models, \cite{[1],[ECM2]}.

Of course while the (OR) and (FD) only hold around equilibrium, {\it
i.e.}  they are properties of $G$--derivatives evaluated at $G=0$, and
the expansion for $\lambda(\beta)$ is a general consequence of the
correlation decay, the (FT) also holds far from equilibrium, {\it i.e.}
for large $G$ and can be considered a generalization of the (OR) and
(FD).

{\it Acknowledgements:} I have profited from many discussions and hints
from F.  Bonetto and P.  Garrido, \cite{[BGG]}.  I am particularly
indebted and grateful to E.G.D.  Cohen for important
comments, criticism, suggestions and much needed encouragement, and
to G. Gentile for pointing out an error in the first
version of the letter and for many comments.  
Partial financial support from Rockefeller U.,
CNR-GNFM, ESI and the EU program: ``Stability and Universality in
Classical Mechanics", \# ERBCHRXCT940460.\*


\begin{references}
\bibitem{[1]} Examples of such thermostats are
the gaussian thermostats and a few concrete cases can be found in
\cite{[GC2]}.  The importance of the gaussian thermostats, and of
reversibility, and the possibility of linking them with the dynamical
systems theory of the SRB distributions,
\cite{[S]}, has been stressed in the fundamental paper \cite{[ECM2]}
whose experimental results and theoretical hints led to the formulation
of the (CH) as a reinterpretation of Ruelle's principle, \cite{[R]}.
\end{references}
\end{document}